\documentclass[a4paper]{spie}  
\def\la{\left\langle}
\def\ra{\right\rangle}
\newcommand{\realR}{{{{\rm I\kern-0.16em{}R}}}}
\newcommand{\mk}{{\mathbf k}}
\newcommand{\mx}{{\mathbf x}}

\usepackage[dvips]{graphicx}

\title{Clustering statistics in cosmology}

\author{Vicent J. Mart\'{\i}nez\supit{a} and Enn Saar\supit{b}
\skiplinehalf
\supit{a}Observatori Astron\`omic, Universitat de Val\`encia, Burjassot, 46100,
Spain \\
\supit{b}Tartu Observatory, T\~oravere, 61602, Estonia
}

\authorinfo{Further author information: (Send correspondence to V.J.M.)
\\V.J.M.: E-mail: Vicent.Martinez@uv.es\\  E.S.: E-mail: saar@aai.ee}

\begin{document}
  \maketitle

\begin{abstract}

The main tools in cosmology for comparing theoretical models with
the observations of the galaxy distribution are statistical. We
will review the applications of spatial statistics to the
description of the large-scale structure of the universe. Special
topics discussed in this talk will be: description of the galaxy
samples, selection effects and biases, correlation functions, Fourier analysis,
nearest neighbor statistics, Minkowski functionals
and structure statistics. Special attention will
be devoted to scaling laws and the use of the lacunarity measures
in the description of the cosmic texture.

\end{abstract}



\section{INTRODUCTION}
\label{sect:intro}  

Cosmology is a science which is experiencing a great development
in the last decades. The achievements in the observations are
driven the subject into an era of precision. The two fundamental
pillars upon which observational cosmology rests are the cosmic
microwave background and the distribution of the galaxies. The
analysis of the huge amount of data that is now being collected in both
areas will provided a unified framework to explain the formation
and evolution of the large-scale structure in the universe. In
this paper we will review some of the aspects related with the
galaxy clustering.

\section{OBSERVATIONS OF THE GALAXY CLUSTERING}
\label{sect:obser}

\subsection{Distances}
\begin{figure}
   \begin{center}
   \begin{tabular}{c}
   \includegraphics[width=6.cm]{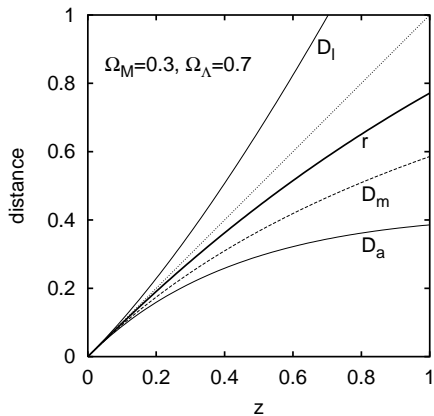}
   \end{tabular}
   \end{center}
   \caption[distances]
   { \label{fig:distances}
Different distance measures for a currently popular 'concordance'
model universe. Here $D_l$ is the luminosity distance, $r$ is the
comoving distance, $D_m$ is the usually used Mattig distance (not defined
for this model), and
$D_a$ is the angular diameter distance. All distances are given in
the units of the Hubble length $c/H_0$.}
   \end{figure}

Astronomers can accurately measure the galaxy positions on the
sky. Unfortunately, it is not possible to have the same accuracy
for the radial distance of each object. Different distance
estimators are used in astronomy (see, for example,
Ref.~\citenum{sgd} for a review). For the luminosity selection
effects one has to use the luminosity distance $D_l$, for the
angular selection effects the angular diameter distance $D_a$, and
in order to describe spatial clustering the comoving distance $r$
is used. All these distances can be derived from the cosmological
redshift of the galaxy $z_{\mbox{\scriptsize cos}}$. Distances however
depend on the adopted cosmological model and the value of its
parameters. For nearby galaxies, the Hubble law states that
$cz_{\mbox{\scriptsize cos}} = H_0 r$ where $H_0$ is the present value
of the Hubble parameter. Recent measurements\cite{Freedman01}
provide a value of $H_0 = 72 \pm 8$ km s$^{-1}$ Mpc$^{-1}$. As
space is curved, for more distant galaxies the distance-redshift
relation is not linear any more, and different distances differ.
We illustrate this in Fig. \ref{fig:distances}, where the
different cosmological distances are given for the presently
popular 'concordance model'. The statistics describing spatial
clustering obviously depend on the adopted distance definitions,
and thus on the prior cosmological model. This should be kept in
mind, as these statistics are frequently used to estimate the
'true' parameters of the cosmological model.

It is important to mention that the true cosmological redshift is
not a measurable quantity since what we really are able to measure
for each galaxy is a quantity $z$ satisfying the relation
$cz=cz_{\mbox{\scriptsize cos}} + v{\mbox{\scriptsize pec}}$ where
$v{\mbox{\scriptsize pec}}$ is the line-of-sight peculiar velocity.
Peculiar velocities create a distorted version of the galaxy
distribution ---namely the redshift space---, as opposed to the real
space where galaxies lie at their real positions. Distortions are
more severe within the high density regions where effects like the
Fingers-of-God ---elongated structures along the line-of-sight---
are the most evident consequence\cite{veldist}. In next sections
we will discuss how this distortions affect the statistical
clustering measures.

\subsection{Recent redshift surveys}
Cosmology, like many other modern scientific branches, is
technology driven.  Nowadays, use of multi-fiber spectrographs
permits to measure the redshifts of many galaxies in a single run.
Numbers have changed from 5-10 redshifts of galaxies measured per
night in the 70s to 2000 redshifts per night at the late 90s. The
reachable magnitude limit has moved from  14 to 19.5 in the
blue\cite{guzzo02}. On the basis of this technology, new huge
surveys of redshifts of galaxies  are being built. The main two
ongoing projects are the 2dF (2-degree Field) and the SDSS (Sloan
Digital Sky Survey). More information about these surveys can be
found in their Web pages: {\tt
http:/\kern-2pt/www.mso.anu.edu.au/2dFGRS/\,} for the 2dF survey
and {\tt http:/\kern-2pt/www.sdss.org/\,} for the SDSS survey.

\begin{figure}
   \begin{center}
   \begin{tabular}{c}
   \includegraphics[height=7.404cm]{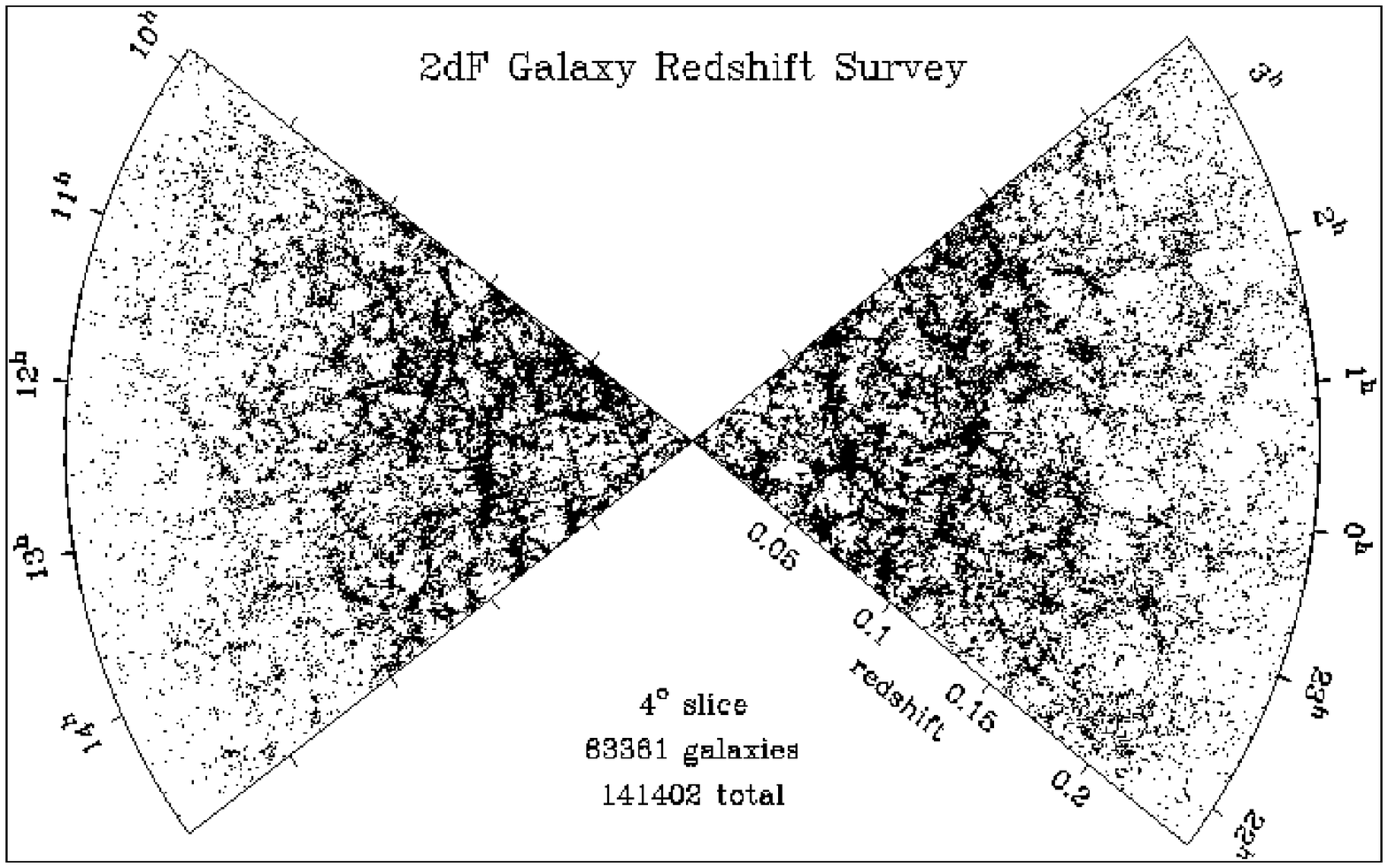}
   \includegraphics[height=1.254cm]{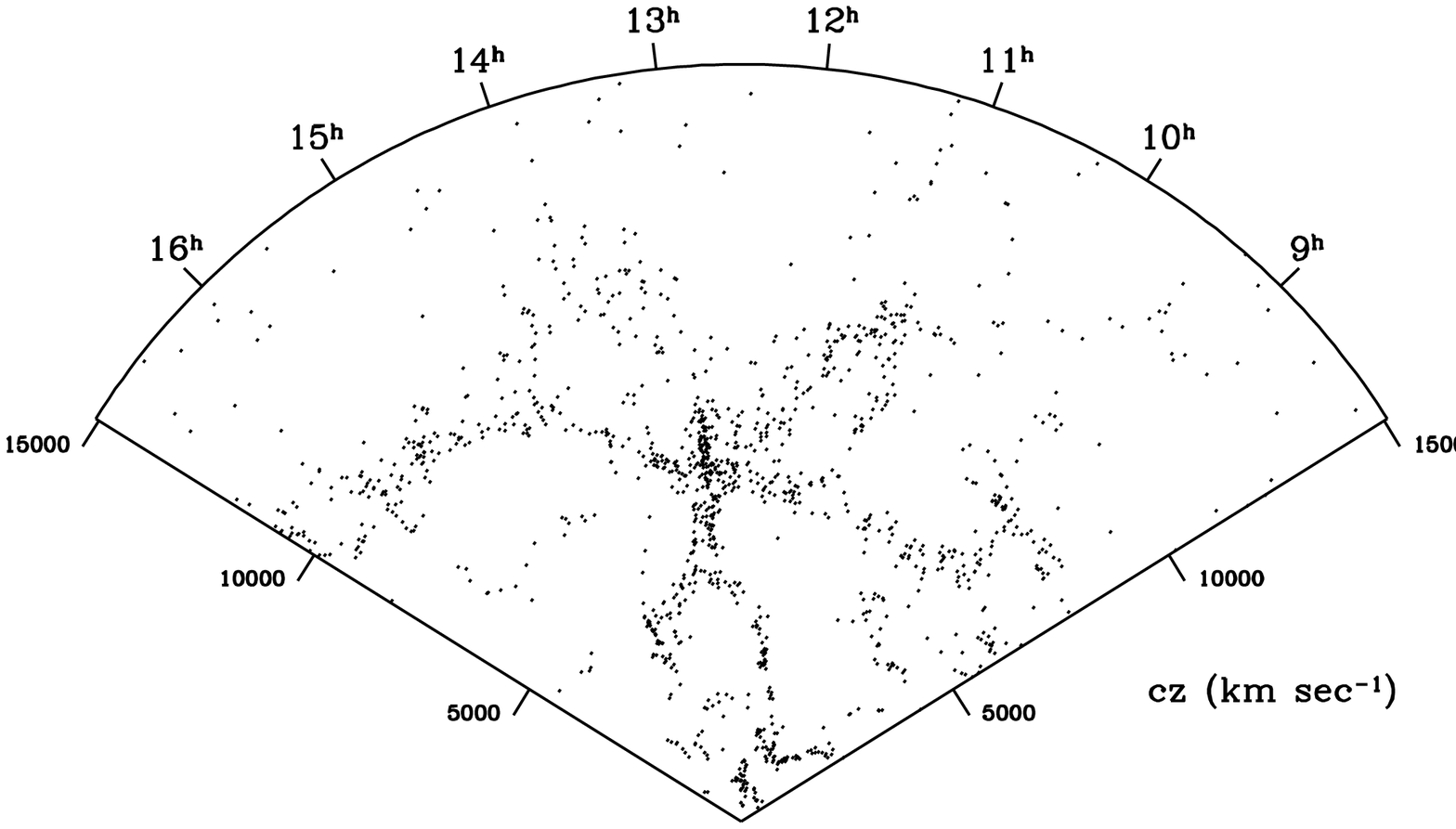} \\
   \includegraphics[height=10.92cm]{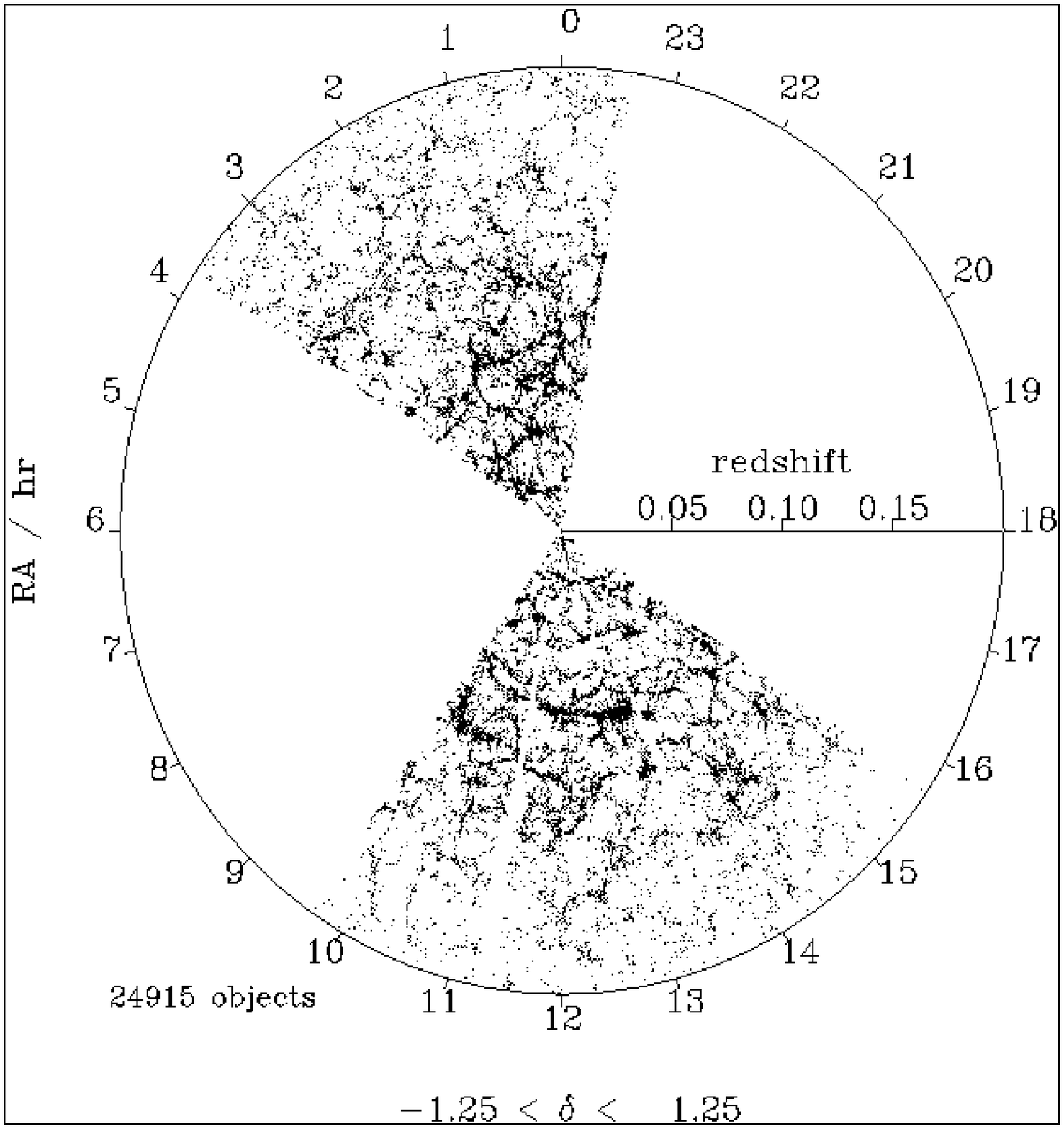}
   \end{tabular}
   \end{center}
   \caption[samples]
   { \label{fig:samples}
The top panel shows two $4^\circ$ slices with 63,381 galaxies from
the 2dF redshift survey. The maximum depth is $z=0.25$, (figure from Peacock
{\it et al.}\cite{2dfnature}). The bottom panel shows two $2.5^\circ$ slices
with 24,915 galaxies from the SDSS survey. The outer circle has
redshift $z=0.2$ (figure from Loveday\cite{loveday}). As an inset, the
first CfA slice from de Lapparent, Geller, and Huchra\cite{lapparent} 
is shown to scale. }
   \end{figure}

In Fig.~\ref{fig:samples} we show cone diagrams of these two
samples under construction. The visual analysis of these plots
reveals the characteristic patterns already noticed in the famous
first slice of the universe\cite{lapparent} (also shown in the
diagram): a bubbly structure in which filaments and walls surround
empty regions nearly devoid of galaxies. The big clusters of
galaxies lie typically in the intersections of this labyrinth of
structures.

Nevertheless, the new surveys display an important difference with
respect to the older shallower slices: The depth of the surveys
---being now several times larger--- has allowed, for the first
time, to be sure that the observed structures are much smaller
than the size of the survey itself. This was not the case at the
end of the 80s when the first Center for Astrophysics (CfA) slice
was compiled. At that time it was not clear if the observed
large-scale structures were going to increase in size with the
survey depth. The first serious evidence on the contrary came with
the Las Campanas Redshift Survey (LCRS)\cite{lcrs,mart99}. This
survey represented the beginning of the end\cite{kir96} in the
sense that the characteristic structures of the galaxy
distribution ---voids, walls, and filaments--- reached a maximum
size and no structures of larger size were observed, as it should
be expected if the distribution of galaxies was to be an unbounded
fractal\cite{mart02}. This trend has been reinforced by the 2dF
and Sloan first cone diagrams (see Fig.~\ref{fig:samples}).

\subsection{Selection effects}

One characteristic feature of these plots is that at larger
distances the number of galaxies decreases. This phenomenon is
entirely due to the way the surveys are built: they are
flux-limited, therefore only the galaxies bright enough to have an
apparent magnitude exceeding the survey cutoff are detected. At
larger distances it is only possible to observe the intrinsically
most luminous galaxies. To account for this incompleteness in the
statistical analysis of these surveys, one needs to know the
selection function $\varphi(r)$, which basically provides the
probability that a galaxy at a given distance $r$ is included in
the sample. This is the radial selection function that is
usually estimated from the ---previously calculated--- luminosity
function, $\phi(L)$. The luminosity function is defined by  
the number density of galaxies in a given range of
intrinsic luminosity $[L,L+dL]$, $\phi(L) dL$.
This function varies with
morphological type, environmental properties and redshift due to
galactic evolution. Traditionally it has been empirically fitted
to a Schechter function\cite{Schechter76}
\begin{equation}
\phi(L) dL = \phi_{\ast} \left ({L \over L_{\ast}} \right )^{\alpha}
\exp \left (- {L \over L_{\ast}} \right ) d \left ( {L \over
L_{\ast}} \right ) ,
\label{Eq:lum}
\end{equation}
where $L_{\ast}$ is a characteristic luminosity which separates
the faint galaxy range where the power-law with exponent $\alpha$
dominates Eq.~\ref{Eq:lum} and the bright end where the number
density decreases exponentially. 

Other selection effects affect the galaxy samples. Many of them are
directional. Some are due to the construction of the sample:
masks in given fields, fiber collisions in the spectrographs,
etc. In addition, the sky is not equally transparent to the
extragalactic light in all directions due to absorption of light
performed by the dust of the Milky Way. Since the shape of our own
Galaxy is rather flat, the more obscured regions are those with
low values of $|b|$ (where $b$ is the galactic latitude). This
effect has to be considered when computing the real brightness of
a galaxy, which therefore depends on the direction of the
line-of-sight. The best way to take this effect into account is to
consider the well defined maps of the distribution of galactic
dust\cite{dust}. 

For very deep samples, the absolute magnitude
has to be estimated considering other adjustments like the K-correction,
which takes into account that the luminosity of the galaxies at
large redshift is detected at longer wavelength than actually was
emitted.

\begin{figure}
   \begin{center}
   \begin{tabular}{c}
   \includegraphics[width=6.5cm]{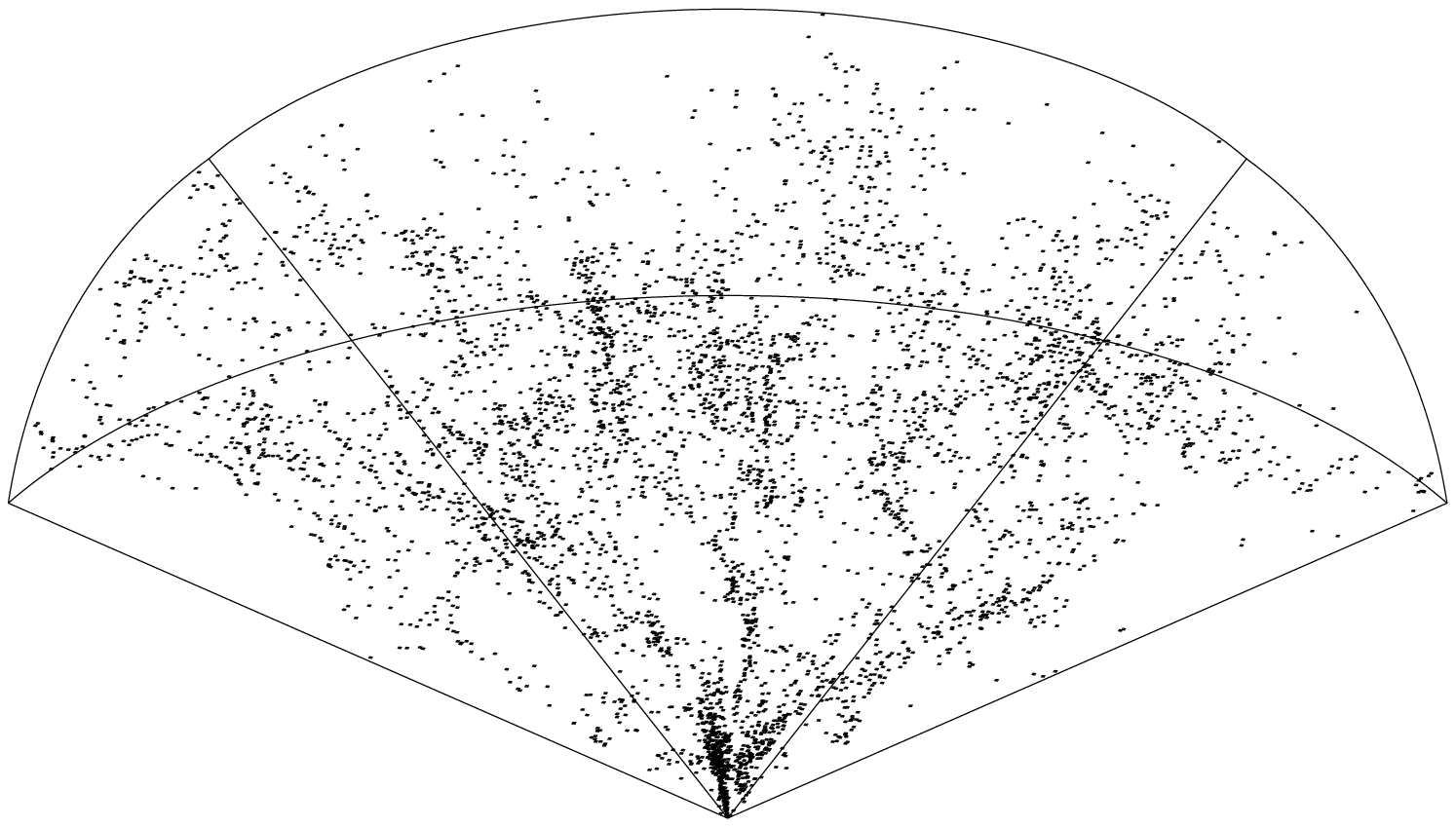}
   \includegraphics[width=6.5cm]{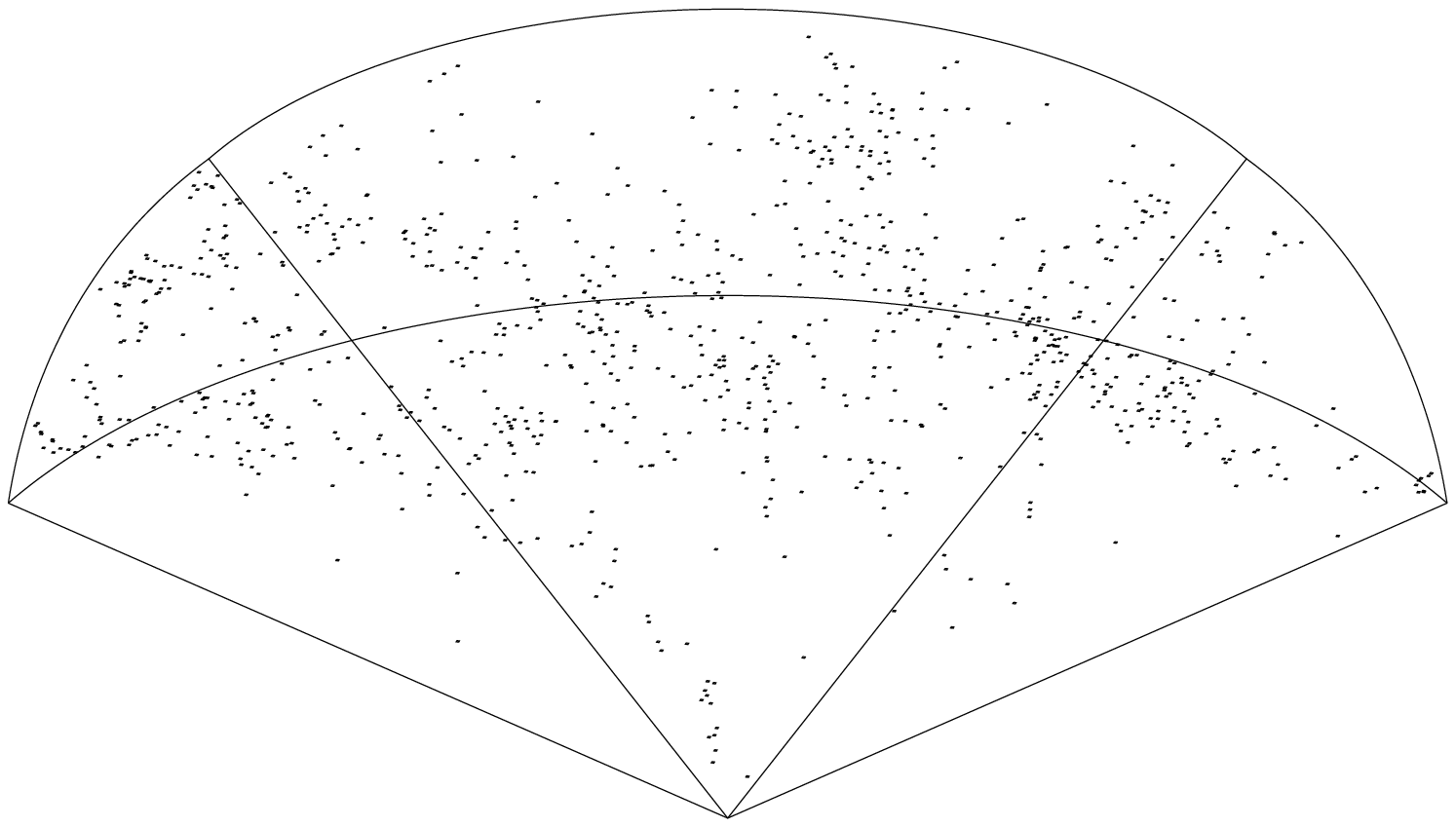}
   \end{tabular}
   \end{center}
   \caption[edge]
   { \label{fig:edge}
   The left panel shows a region in the northern hemisphere of the CfA
apparent magnitude limited sample. The apparent magnitude cutoff is
$m=15.5$. It contains nearly 5000 galaxies. The right panel shows
the same region when considered volume-limited sample. Now only
the 905 galaxies brighter than $-19.70 + 5\log h$ has been
included. The drop of the galaxy density with the distance to the
observer (located at the vertex) appreciated in the left panel is
not longer observed in the volume-limited sample. }
   \end{figure}

Once the selection effects have been considered, the statistical
analysis of the galaxy surveys are performed assigning to each
galaxy a weight inversely proportional to the probability that
the galaxy was included in the sample. A more clean solution is to
consider volume-limited samples at the price of throwing away a
huge amount of the collected data: At a given distance limit, one
can easily calculate the absolute magnitude of a galaxy having the
apparent magnitude limit of the survey. All galaxies intrinsically
fainter that this absolute magnitude cutoff will be ignored in the
volume-limited sample. An illustration of this procedure is shown
in Fig.~\ref{fig:edge}.

\section{CORRELATIONS}

\subsection{The two-point correlation function}

The structure of the universe qualitatively described in the
previous section needs to be quantified by means of statistical
measures having the capacity of distinguish between different
point patterns.

The most popular measure used in this context has been the
two-point correlation function\cite{peeb80,sgd} $\xi(r)$. The
first time this quantity was applied to a galaxy catalog was in
1969 by Totsuji and Kihara\cite{totsuji}. Since then, its use has
been widely spread. The quantity $\xi(r)$ is defined in terms of the
probability that a galaxy is observed within a volume $dV$ lying
at a distance $r$ from an arbitrary chosen galaxy,
\begin{equation}
dP = n [1+\xi(r)] dV,
\end{equation}
where $n$ is the average galaxy number density. For a completely
random distribution $\xi(r)=0$. Positive values indicate
clustering, negative values indicate anti-clustering or
regularity. In this definition, isotropy and homogeneity of the
point process is being assumed, otherwise the function $\xi(r)$
should depend on a vector quantity.

Several estimators have been used to obtain the two-point
correlation function from a given data set\cite{pons,kerm}. At
short distances their results are nearly indistinguishable; at
large distances, however, the differences become important. The
best performance is reached by the Hamilton\cite{ham93} and the
Landy and Szalay\cite{landy93} estimators.

   \begin{figure}
   \begin{center}
   \begin{tabular}{c}
   \includegraphics[width=5.cm]{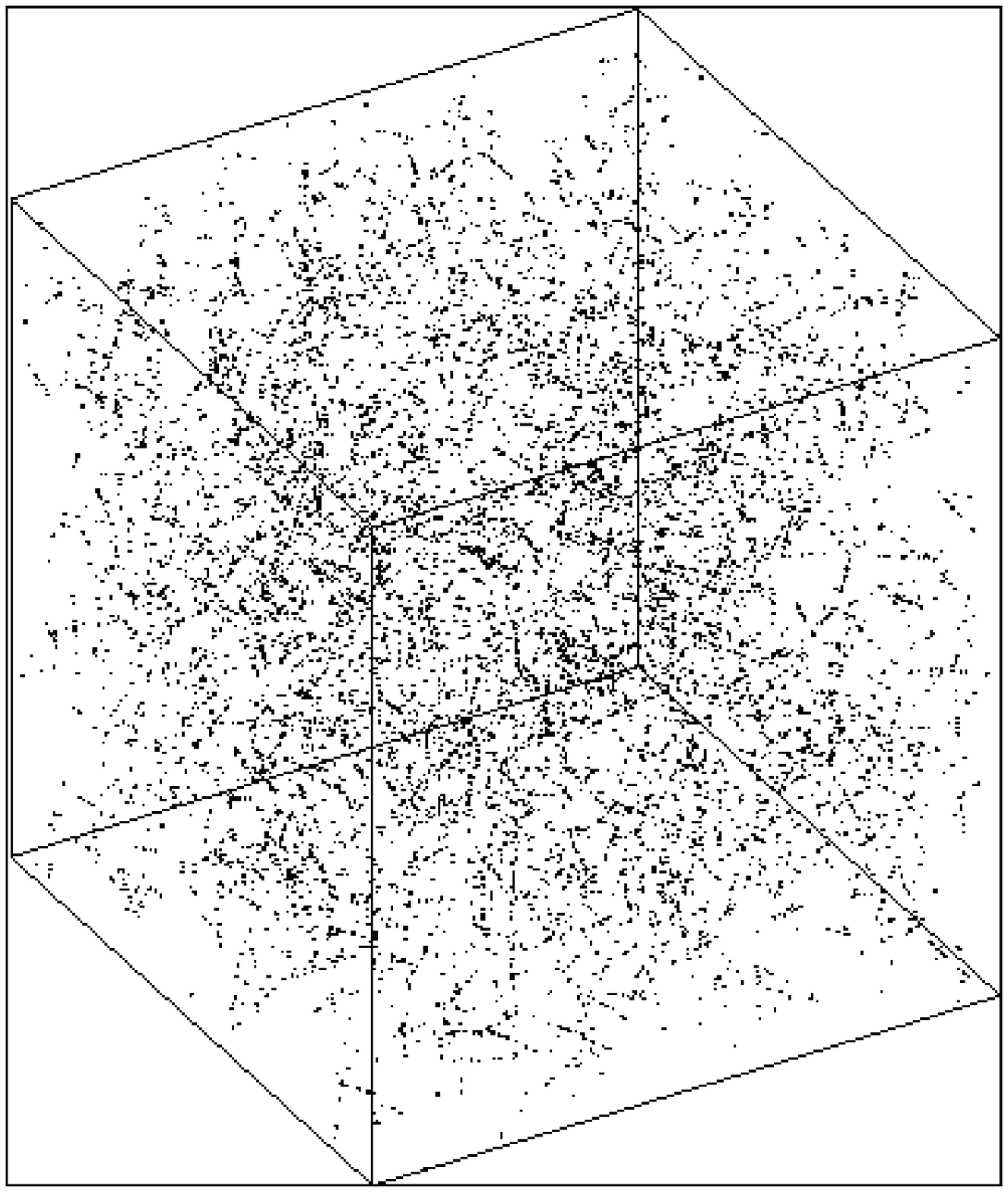}
   \includegraphics[width=8.5cm]{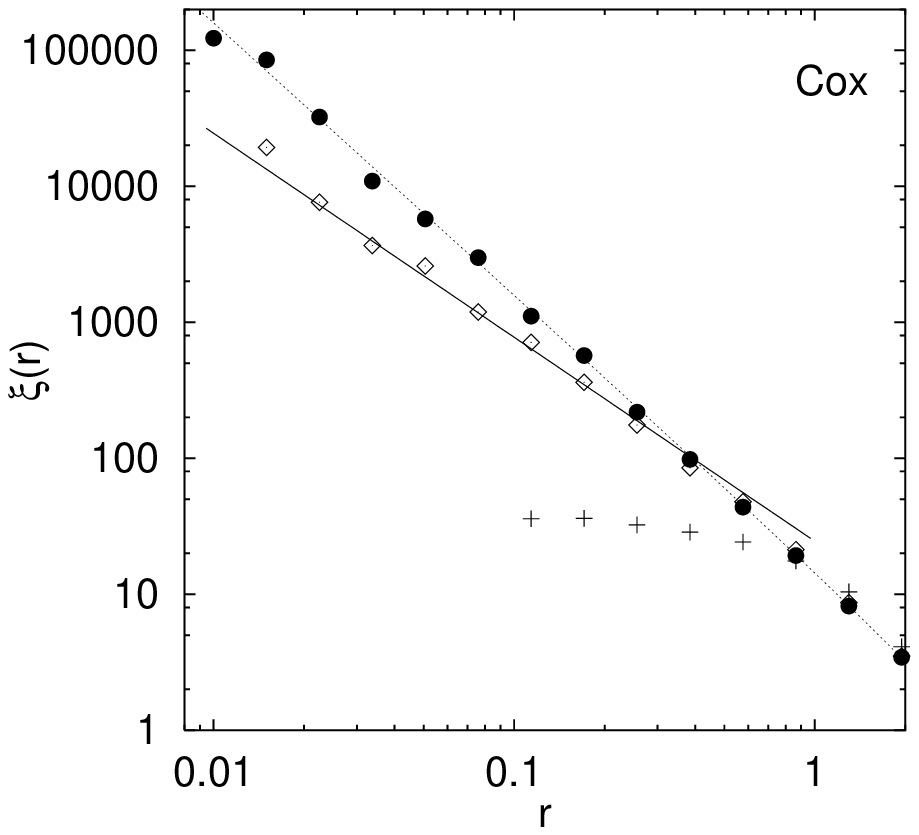}\\
   \includegraphics[width=5.cm]{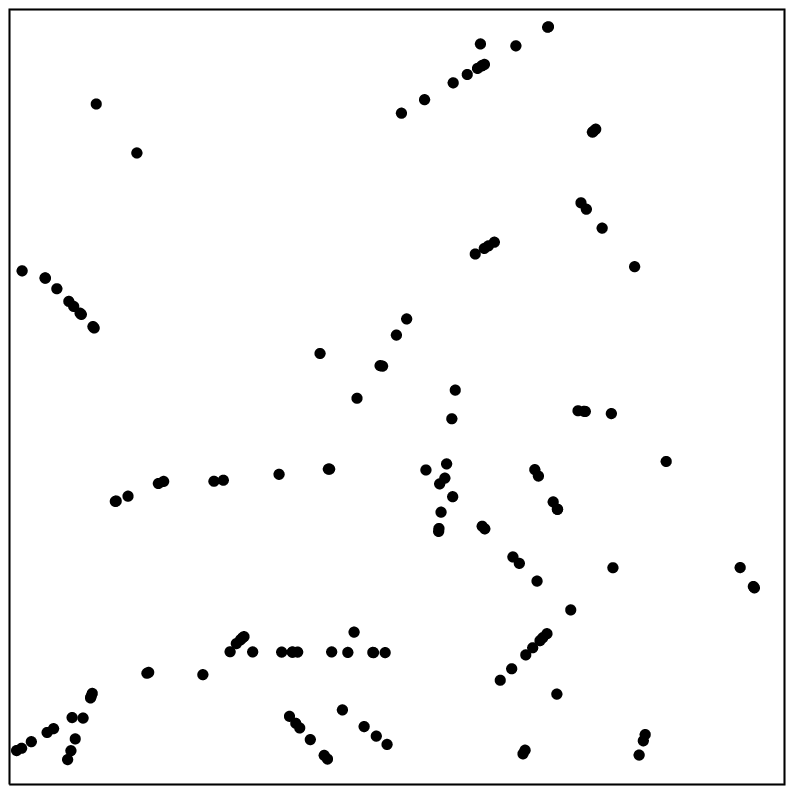}
   \includegraphics[width=5.cm]{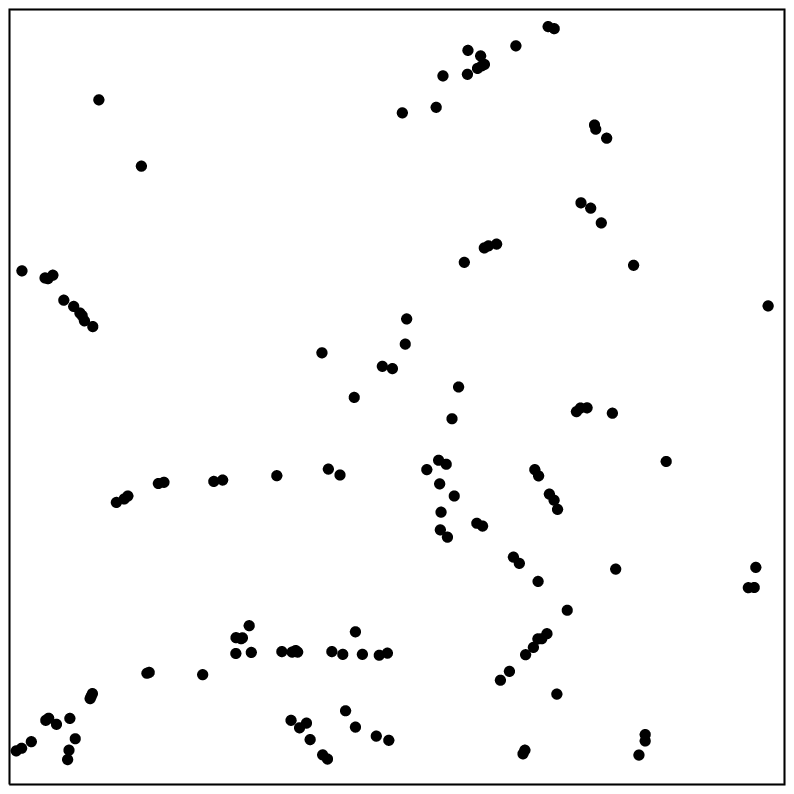}
  \includegraphics[width=5.cm]{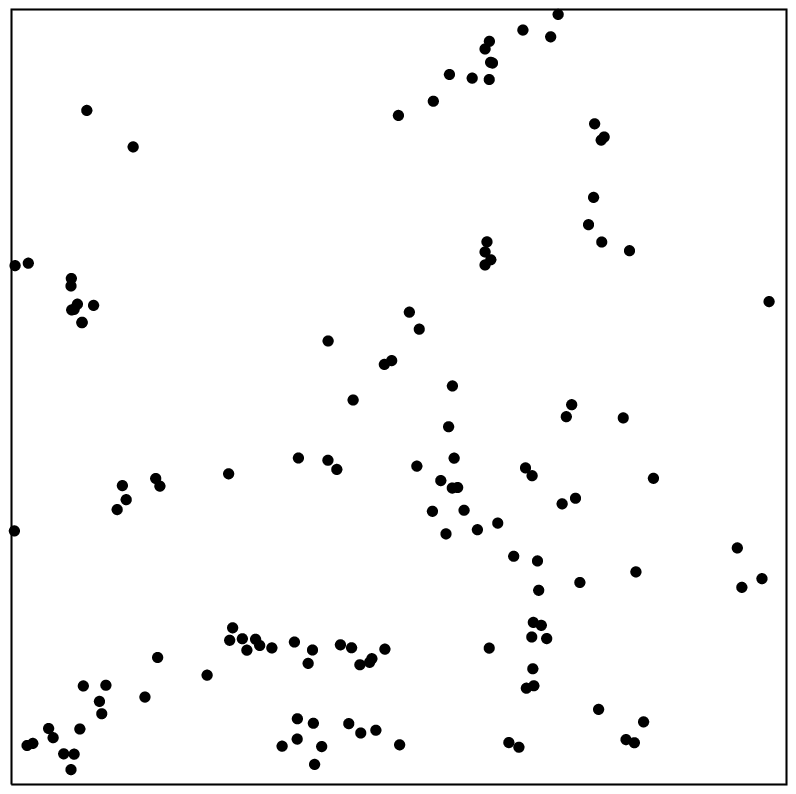}
   \end{tabular}
   \end{center}
   \caption[Coxfig]
   { \label{fig:Coxfig}
The top left panel shows a segment Cox process simulated on a cube
with side-length 100. The top right panel shows
the two-point correlation function: The dotted line
corresponds to the expected analytical expression (see
Eq.~\ref{eq:skmfor}), that is, in this range of scales, close to a power-law
with exponent $-2$. Solid bullets are the empirically calculated
values of $\xi(r)$. Open diamonds correspond to the function
calculated on the shifted point process with shifts following a
power-law distribution function, while crosses correspond to
Gaussian shifts. The left bottom panel shows a slice with dimensions
$40\times40\times10$ drawn from the unshifted full realization. The
same slice is shown after applying power-law shifts (central bottom
panel) and Gaussian fits (right bottom panel).}
   \end{figure}

To illustrate the kind of information that we can extract from
this second-order spatial statistic we can use a point process
having an analytic expression for its two-point correlation
function. A segment Cox process is generated by randomly placing
segments of length $\ell$ within a window $W$. Then, we scatter
points on the segments with a given intensity. If the mean number
of segments per unit volume is $\lambda_s$, the correlation
function of the process has the form\cite{skm}
\begin{equation}
\xi_{\rm {Cox}}(r)=\frac{1}{2\pi r^2\lambda_s \ell}-\frac{1}{2\pi r
\ell^2 \lambda_s},
\label{eq:skmfor}
\end{equation}
for $r \le \ell$ and vanishes for larger $r$. Note that this
expression is independent of the number of points per unit length
scattered on each segment.

In Fig.~\ref{fig:Coxfig} we show a 3-D simulation of this process
with parameters $\lambda_s=0.001$ and $\ell=10$. The correlation
function estimate is shown together with the analytical
expectation of Eq.~\ref{eq:skmfor}. Note that, at small scales,
$\xi_{\mbox{\scriptsize Cox}}(r) \sim r^{-\gamma}$ with $\gamma =2$.  
The strong clustering signal
of this point field can be smeared out by applying independent
random shifts to each point of the simulation. If the random
shifts are performed by a three-dimensional Gaussian distributed
vector with $\sigma=0.5$, the short scale correlations are completely
destroyed (see Fig.~\ref{fig:Coxfig}). 
If the shifts are distributed according to a power-law
density probability function $d^\ast(r) \propto r^\alpha$, the
value of $\gamma$ is
reduced by $2(1+\alpha)$. In the example $\alpha=-0.75$ and therefore
$\gamma$ changes from $2$ to $1.5$. This
seems to be a rather general phenomenon\cite{snethlage}. The
random shifts affect the correlation
function mimicking the way peculiar velocities suppress 
the short range correlations\cite{mart93} (for
scales $r \leq 2 h^{-1}$ Mpc, where $h$ is the Hubble parameter in units
of 100 km s$^{-1}$ Mpc$^{-1}$).

\subsection{$\xi(r)$ on recent samples}

\begin{figure}
   \begin{center}
   \begin{tabular}{c}
\resizebox{0.55\textwidth}{!}{\includegraphics*{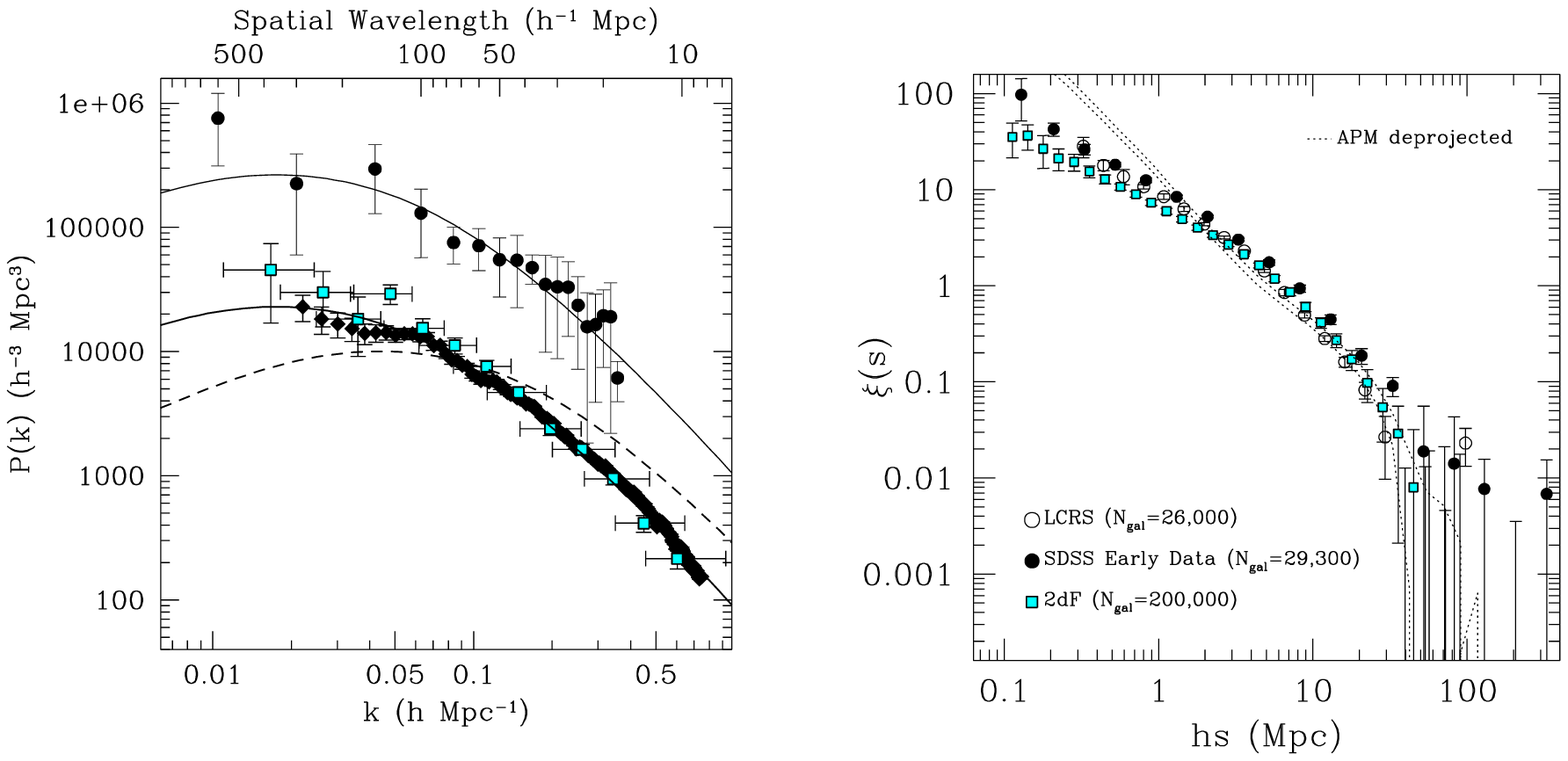}}
     \end{tabular}
   \end{center}
   \caption[xirecent]
   { \label{fig:xirecent}
The two-point redshift correlation function for the deepest
available redshift surveys: The Las Campanas Redshift Survey\cite{Tucker}, the
2dF (Hawkins et al. ---the
2dFGRS team---, in preparation), and the early public release of 
the Sloan Digital Sky Survey\cite{Zehavi}. The
dotted lines, that fit well a power law, correspond to the
real-space correlation function deprojected from the APM angular
data using two different models of galactic evolution\cite{baugh}, 
(figure from Guzzo\cite{guzzo02}).}
   \end{figure}
Fig.  \ref{fig:xirecent} shows the two-point correlation function
in redshift space calculated on three different deep redshift
surveys: LCRS\cite{Tucker}, d2F (Hawkins et al. ---the
2dFGRS team---, in preparation), and SDSS\cite{Zehavi}. The
agreement is quite remarkable, in fact the differences between 
LCRS and SDSS are mainly due to the fact that comoving distances have
been calculated assuming different cosmological models. It is
clear that trying to fit reasonably well
a power-law $\xi(s)=(s/s_0)^{-\gamma}$ to the data is hopeless. In
fact, the dotted lines show the real space correlation function
calculated from the Automatic Plate Machine (APM) angular
data\cite{baugh} after deprojecting the angular correlation
function with the Limber equation. Now, a reliable power-law
$\xi(r)=(r/r_0)^{-\gamma}$, with $\gamma=1.7$ and $r_0=4.1
\,h^{-1}$ Mpc, can be fitted to the curves for scales $ r \leq 4
\,h^{-1}$ Mpc. The slope is in agreement with the results inferred
by Zehavi et al.\cite{Zehavi} for the SDSS early data:
$\gamma=1.75 \pm 0.03$ and $r_0=6.1 \pm 0.2 \, h^{-1}$ Mpc, within
the range $0.1 \le r \le 16 \, h^{-1}$ Mpc.  Although the APM
amplitude was smaller, Baugh\cite{baugh} reported an appreciable
shoulder in $\xi(r)$ for scales $4 \leq r  \leq 25 \,h^{-1}$ Mpc
where the correlation function was rising above the fitted power
law. The diagram also illustrates the effects of the peculiar
velocities in redshift surveys suppressing the short-range
correlations and enhancing the amplitude at intermediate scales
due to coherent flows \cite{veldist,guzzo02}. It is also
interesting to note that the first zero crossing of the two-point
correlation function occurs at scales around 30--40 $h^{-1}$ Mpc.

\subsection{Fractal scaling}

The number of neighbors ---on average--- a galaxy has within a sphere of
radius $r$ is just the integral
\begin{equation}
N(<r) =n \int_0^r 4 \pi s^2 (1+\xi(s)) ds
\label{corrint}
\end{equation}
When this function follows a power-law $N(<r) \propto r^{D_2}$,
the exponent is the correlation dimension, and the point pattern
is said to verify fractal scaling. There is no doubt that up to a
given scale the galaxy distribution fits rather well the fractal
picture. However, some controversy regarding the extent of the
fractal regime has motivated an interesting
debate\cite{pietronero,davisprin,guzzo,mart99}. Nevertheless, the
new data is showing overwhelming evidence that the correlation
dimension is a scale dependent quantity. Different authors
\cite{mart98,cappi98,wuetal,scaramel,hatton,amend,pancoles,lcfrac,colless}
have analyzed the more recent available redshift surveys using
$N(r)$ or related measures with appropriate estimators. Their
results show unambiguously an increasing trend of $D_2$ with the
scale from values of $D_2 \sim 2$ at intermediate scales to values
of $D_2 \sim 3$ at scales larger than 30 $h^{-1}$ Mpc. Moreover,
one of the strong predictions of the fractal hypothesis is that
the correlation length ---the value $r_0$ at which the correlation
function reaches the unity ($\xi(r_0)=1$)--- must increase linearly
with the depth of the sample. This seems not to be the
case\cite{davis88,mart01}.

\subsection{Lacunarity}
If in a fractal distribution, we count for each point the total
number of neighbors within a ball of radius $r$, $M(r)$, we can
see that this quantity follows roughly a power-law
\begin{equation}
M(r)= F r^D,\label{mrr}
\end{equation}
the exponent $D$ is the so-called mass-radius dimension. Taking
the average over all the points we get an estimate of the integral
correlation function $N(<r)=\la M(r) \ra$.  In  this section, we
show how the correlation dimension alone is not enough to
characterize the fractal structure.

The variability of the prefactor $F$ in Eq. \ref{mrr} can be used
as a measure to distinguish between different fractal patterns
having the same correlation dimension. This variability provides an
indicator of the lacunarity. Several alternative quantitative
measures have been proposed in the
literature\cite{Allain,Plotnick,Gaite}. According to
Ref.~\citenum{blum97}, we adopt the following numerical definition
for the lacunarity, which is basically the second-order
variability measure of the prefactor $F$ in Eq.~\ref{mrr},
\begin{equation}
\Phi =\frac{\la (F -\la F \ra )^2 \ra}{\la F \ra ^2}. \label{lac}
\end{equation}

\begin{figure}
   \begin{center}
   \begin{tabular}{c}

(a)   \includegraphics[width=3.5cm]{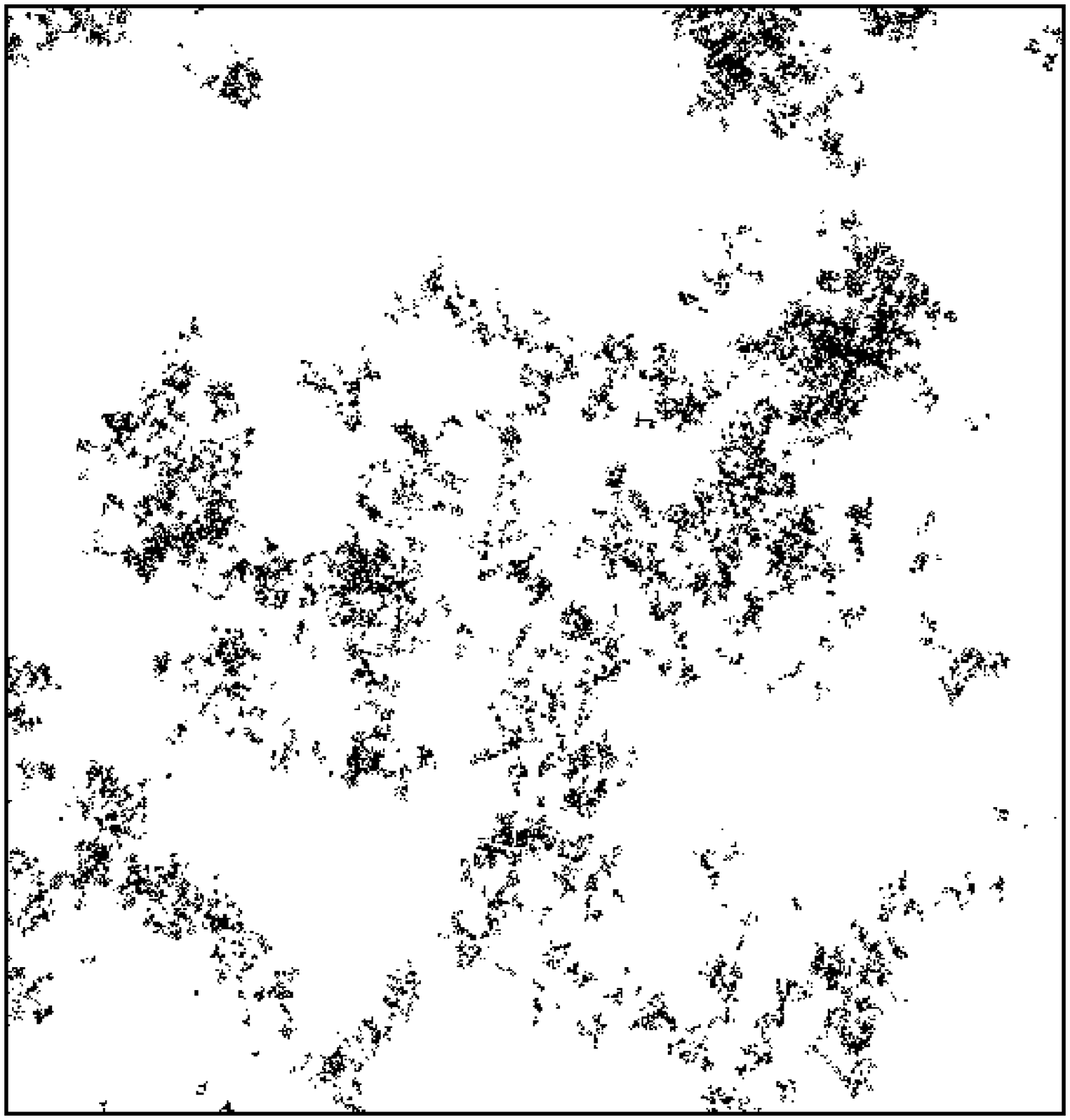}
(b)   \includegraphics[width=3.5cm]{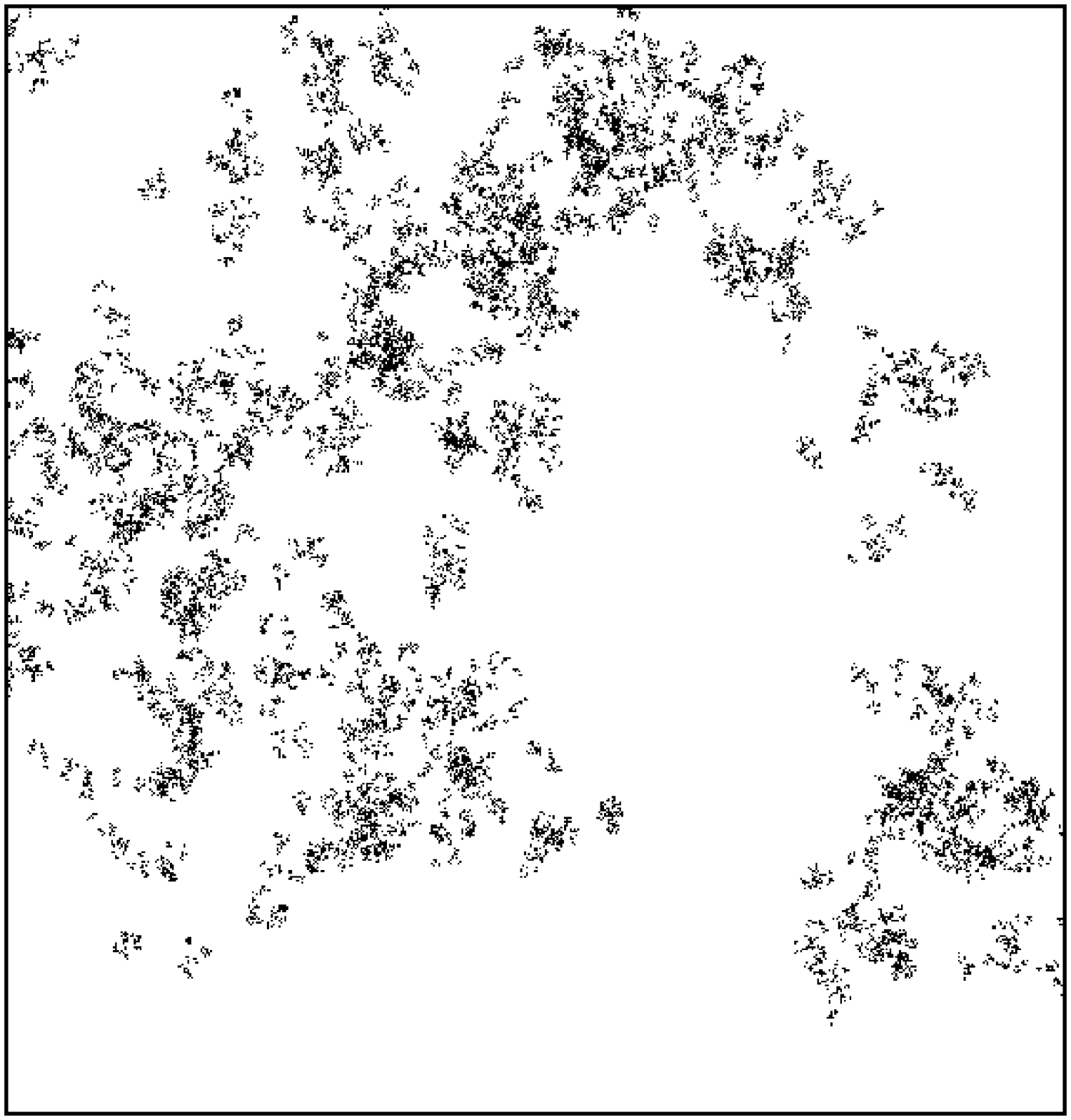}
(c) \includegraphics[width=3.5cm]{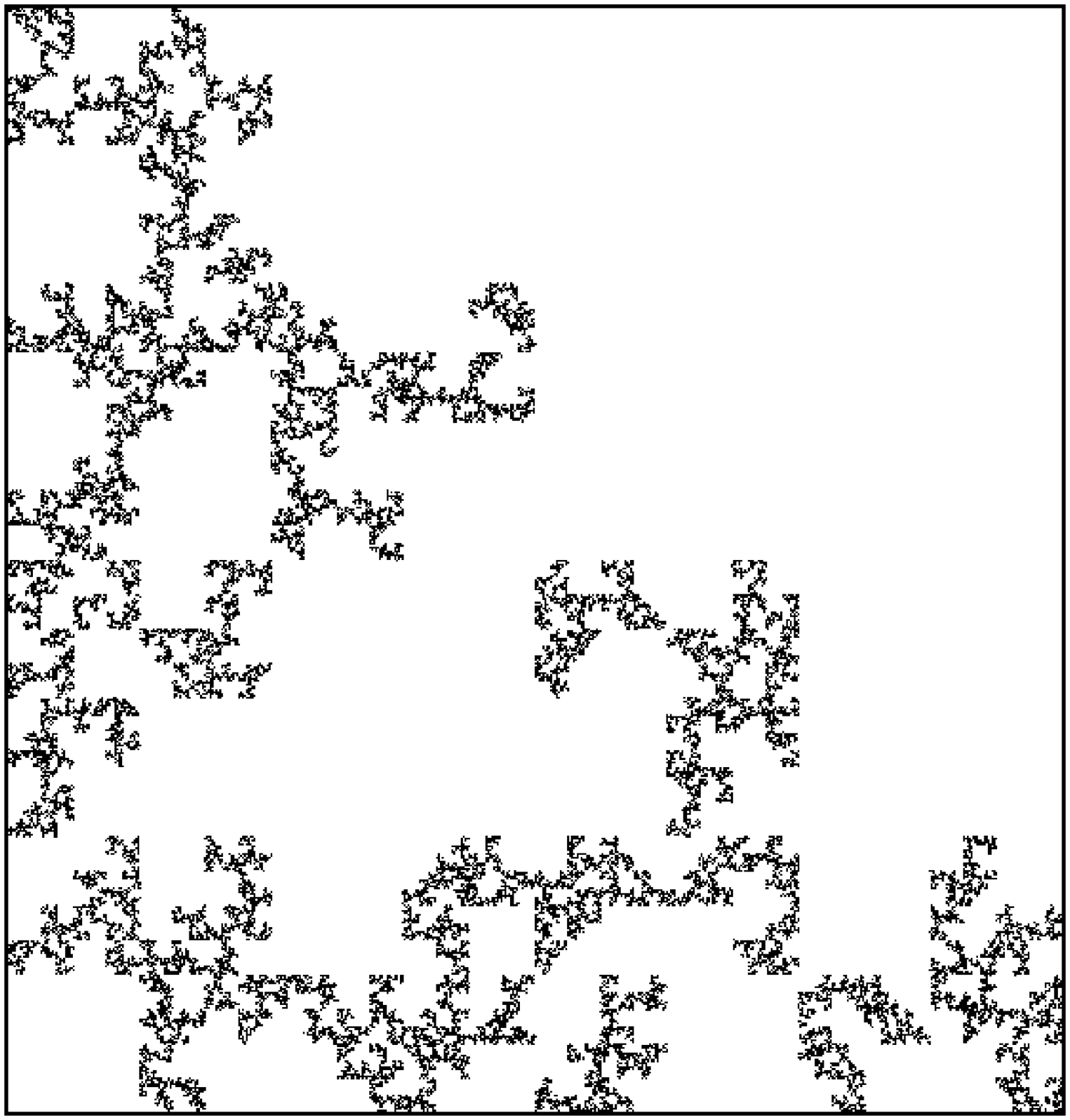}
(d)   \includegraphics[width=3.5cm]{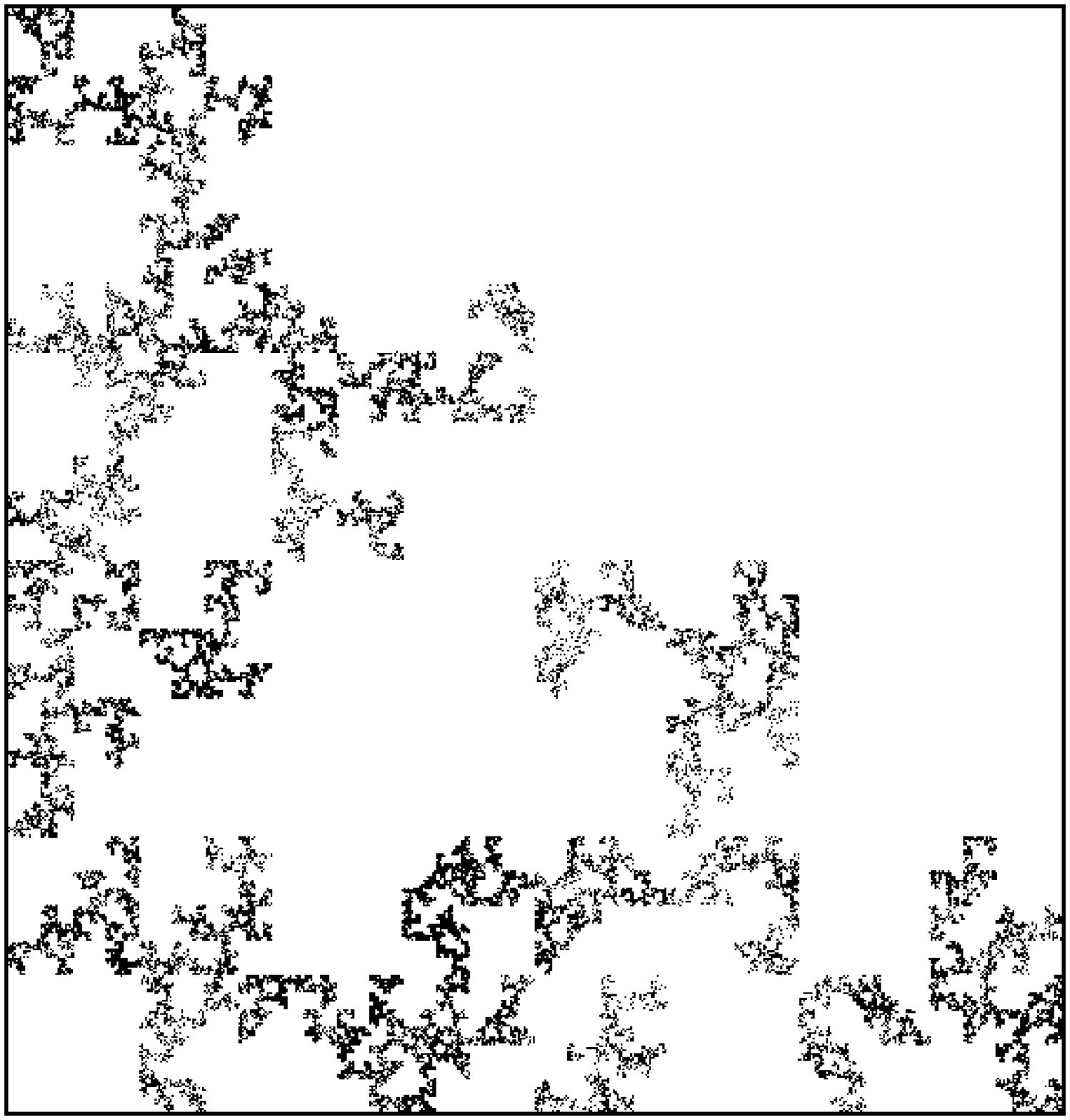} \\
(e)
   \includegraphics[width=10.cm]{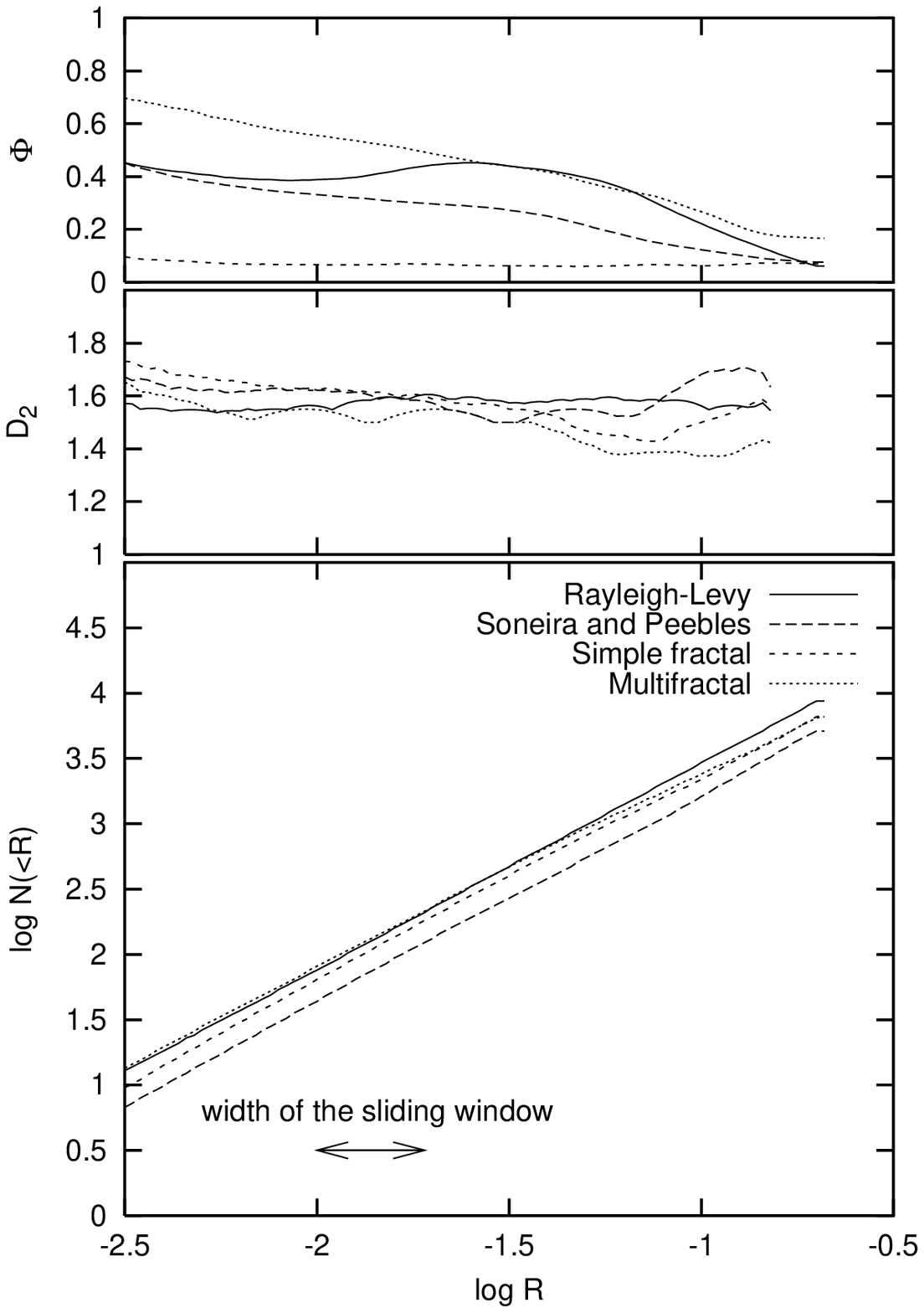}

   \end{tabular}
   \end{center}
   \caption[lac]
   { \label{fig:lac}Panels (a)-(d) show different point patterns having
similar correlation dimension: (a) Rayleigh-L\'evy dust, (b)
Soneira and Peebles model, (c) simple fractal distribution, (d)
multifractal distribution. For all cases, the correlation integral follows
a power law with similar exponent (panel (e) bottom). The
central diagram in panel (e)
shows the local correlation dimension calculated as the slope
of the log-log linear regression within small portions 
of the scale range ---the width of the sliding window
is displayed by an arrow---. The lacunarity measure
characterizing the textural properties of each point pattern is
shown in the top diagram of panel (e).
 }
   \end{figure}

We first illustrate these measure on several two-dimensional point
patterns.
\begin{enumerate}
\item
Mandelbrot\cite{mand75} proposed an elegant fractal prescription
to locate galaxies in space. It is the so-called Rayleigh-L\'evy
flight: galaxies are placed at the end points of a random walk
with steps having isotropically random directions. The step length
follows a power-law probability distribution function
$P(r>\ell)=(\ell_0/\ell)^D$ for $\ell \ge \ell_0$ with $D<2$, and
$P(r>\ell)=1$ for $\ell < \ell_0$. Panel (a) in Fig. \ref{fig:lac}
shows a two-dimensional simulation with $D=1.5$ and $\ell=0.001$
generated within a square with sidelength 1.

\item
Soneira and Peebles\cite{sp} proposed a hierarchical fractal model
to mimic the statistical properties of the Lick galaxy catalog.
This model is built as follows: Within a sphere of radius $R$ we
place randomly $\eta$ spheres of radius $R/\lambda$ with
$\lambda>1$. Now, in each of the new spheres, $\eta$ centers of
smaller spheres with radius $R/\lambda^2$ are placed. This process
is repeated until a given level $L$ is reached. Galaxies are
situated at the centers of the $\eta^L$ spheres of the last level.
The correlation dimension of this fractal clump is $\log (\eta)
/\log (\lambda)$. Panel (b) in Fig. \ref{fig:lac} shows a
two-dimensional simulation with $\eta=2$, $\lambda=1.587$, and
therefore $D_2=1.5$. Four clumps with $L=13$ have been generated
within a disc of diameter 1.

\item
The next example is a multiplicative cascade process performed on
the unit square\cite{mart90}. First, the square is divided into four
equal pieces. We assign a probability measure to each of the pieces
randomly permuted from the set $\{p_1,p_2,p_3,p_4\}$. Each
subsquare is divided again into four pieces, and again we attach a
measure to each of the them by multiplying a $p_i$ value randomly
permuted by the value corresponding to its parent square. The
process is repeated several times, and in each step the measure
attached to each small square is the product of a new $p_i$ value
with all its ancestors. After $L$ steps, we end with a mass
distribution over a $2^L\times 2^L$ lattice. A point process is
then generated placing randomly points within each pixel with 
probability proportional to its measure. Panels (c) and (d) in
Fig. \ref{fig:lac} show two realizations of this model, one being
a simple fractal, panel (c), with $p_1=p_2=p_3=1/3$, and $p_4=0$  and
the other one being a multifractal measure, panel (d), with $p_1= 0.4463,\,\;
p_2=0.2537,\,\; p_3=0.3$, and $p_4=0$.
While for the first case $D_2 = \log 3/\log 2 \simeq 1.58$, for
the multifractal measure the chosen values of $p_i$ provide a
dimensionality $D_2 = 1.5$.

\end{enumerate}

The bottom panel of Fig. \ref{fig:lac} shows the relation $\log
N(<r)$ versus $\log r$ for the four examples. The power-law
behavior $N(<r) \propto r^{D_2}$ is clearly appreciated in the
diagram, with scaling exponent $D_2 \simeq 1.6$ for all cases.

A more detailed analysis of the {\it local} correlation dimension
is reported in the central diagram of panel (e),
where we show how $D_2$ changes with the
scale. In this case, $D_2$ has been calculated as the slope of the
local linear regression fit to a small portion of the curve. This
sliding window estimate of the local value of $D_2$ is very
sensitive to any possible non fractal behavior that could not well
be appreciated in the plot of $\log N(<r)$. The width of the
sliding window used in the estimation is shown as an arrow in
the bottom panel. We can see that in all the analyzed point patterns the
empirical local correlation dimension oscillates around $D_2
\simeq 1.6$. It is therefore rather hard to find significative
differences between the analyzed patterns through the function
$N(<r)$ or from $D_2(r)$.

The differences, however, are revealed by the lacunarity measure
(Eq. \ref{lac}) which is shown in the top diagram of panel (e).
The lacunarity curves $\Phi(r)$, associated to each pattern, show
clear differences between them providing us with a valuable
information about the texture of each process.

The simple fractal model in panel (c) shows rather constant
behavior of $\Phi$ with the scale, displaying only very small
oscillations around $\Phi \simeq 0.1$. By contrast, the
multifractal set, being quite similar to the eye to the simple
fractal, shows a completely different lacunarity curve, with a
characteristic monotonic decreasing behavior from $\Phi \simeq
0.7$, at the smallest scales, to $\Phi \simeq 0.2$ at the larger
scales. In this case the lacunarity is associated to the
inhomogeneous distribution of the measure on the fractal
support\cite{Allain,Plotnick} in which we can find highly populated
regions (where the values of the measure are very large) together
with other nearly empty locations (where the measure takes the
lowest values). The lacunarity measure reveals the small scale
heterogeneity of the multifractal set. Only at large scales the
curve approaches that of the simple fractal pattern.

The lacunarity curve of the Soneira and Peebles model (panel (b))
is quite similar to that of the multifractal cascade model, with a
decreasing behavior of $\Phi$ with the scale. We can see in the
plot that $\Phi$ varies from $0.45$ at the smallest scales to
$0.08$ at the largest analyzed distances. Because the different
clumps overlap with each other, the set presents scale-dependent
structure which cannot be discovered by analyzing the correlation
function or the correlation dimension alone.

It is quite remarkable how the lacunarity curve of this model
differs from the one corresponding to the Rayleigh--L\'evy flight,
although both spatial patterns seem quite similar to the eye.
Within the first 2/3 of the analyzed scale range, the behavior of
$\Phi$ with the scale, for the Rayleigh--L\'evy dust, is rather
flat with oscillations around $\Phi \simeq 0.4$. It is, therefore,
qualitatively similar to the behavior of the simple fractal
pattern, although showing a higher value of $\Phi$ and displaying
oscillations with higher amplitude. The large-scale properties of
finite regions of Rayleigh--L\'evy dusts are extremely variable,
and the rapid decrease of lacunarity at larger scales for the
sample shown in panel (a) is typical only for dense subregions of
a Rayleigh--L\'evy flight.

\section{POWER SPECTRUM}

The power spectrum (power spectral density) is also a quadratic
statistic of the spatial clustering, as is the two-point
correlation function. Formally they are equivalent (the power
spectrum is the Fourier transform of the correlation function),
but they describe different sides of a process. The power spectrum
is more intuitive physically, separating processes on different
scales. Moreover, the model predictions are made in terms of power
spectra. Statistically, the advantage is that the power spectrum
amplitudes for different wavenumbers are statistically orthogonal:
\begin{equation}
E\left\{\widetilde{\delta}(\mk)\widetilde{\delta}^\star(\mk')\right\}=
    (2\pi)^3\delta_D(\mk-\mk')P(\mk).
\end{equation}
Here $\widetilde{\delta}(\mk)$ is the Fourier amplitude of the
overdensity field $\delta=(\rho-\bar{\rho})/\bar{\rho}$ at a
wavenumber $\mk$, $\rho$ is the matter density, a star denotes
complex conjugation, $E\{\}$ denotes expectation values over
realizations of the random field, and $\delta_D(\mx)$ is the
three-dimensional Dirac delta function.

If we have a sample (catalog) of galaxies with the
coordinates $\mx_j$, we can write the estimator for a Fourier amplitude
of the overdensity distribution\cite{fkp}
(for a finite set of frequencies $\mk_i$) as
\[
F(\mk_i)=\sum_j\frac{\psi(\mx_j)}
    {\bar{n}({\mx}_j)}e^{i\mk_i\cdot\mx} -\widetilde{\psi}(\mk_i),
\]
where $\bar{n}(\mx)$ is the position-dependent selection
function (the observed mean number density) of the sample and
$\psi(\mx)$ is a weight function that can be selected
at will.

The raw estimator for the spectrum is
\[
P_R(\mk_i)=F(\mk_i)F^\star(\mk_i),
\]
and its expectation value
\[
E\left\{\langle|F(\mk_i)|^2\rangle\right\}
    =\int G(\mk_i-\mk')P(\mk')\,\frac{d^3k'}{(2\pi)^3}
    +\int_V\frac{\psi^2(\mx)}{\bar{n}(\mx)}\,d^3x,
\]
where $G(\mk)=|\tilde{\psi}(\mk)|^2$ is the window
function that also depends on the geometry of the
sample volume.
Symbolically, we can get the estimate of the
power spectra $\widehat{P}$ by inverting the integral equation
\[
G\otimes \widehat{P}=P_R-N,
\]
where $\otimes$ denotes convolution,
$P_R$ is the raw
estimate of power, and $N$ is the (constant) shot noise term.

    In general, we have to deconvolve the noise-corrected
raw power to get the estimate of the power spectrum.
A sample of a characteristic spatial
size $L$ creates a window function of width of $\Delta k\approx 1/L$,
correlating estimates of spectra at that wavenumber interval.

As the cosmological spectra are usually assumed to be isotropic,
the standard method to estimate the spectrum involves an
additional step of averaging the estimates $\widehat{P}(\mk)$ over
a spherical shell $k\in[k_i,k_{i+1}]$ of thickness $k_{i+1}-k_i>
\Delta k=1/L$ in  wavenumber space.

As the data set get large, straight application of direct methods
(especially the error analysis) becomes difficult. There are
different recipes that have been developed with the future data
sets in mind. A good review of these methods is given in
Ref.~\citenum{future}.

\begin{figure}
   \begin{center}
   \begin{tabular}{c}
\resizebox{0.55\textwidth}{!}{\includegraphics*{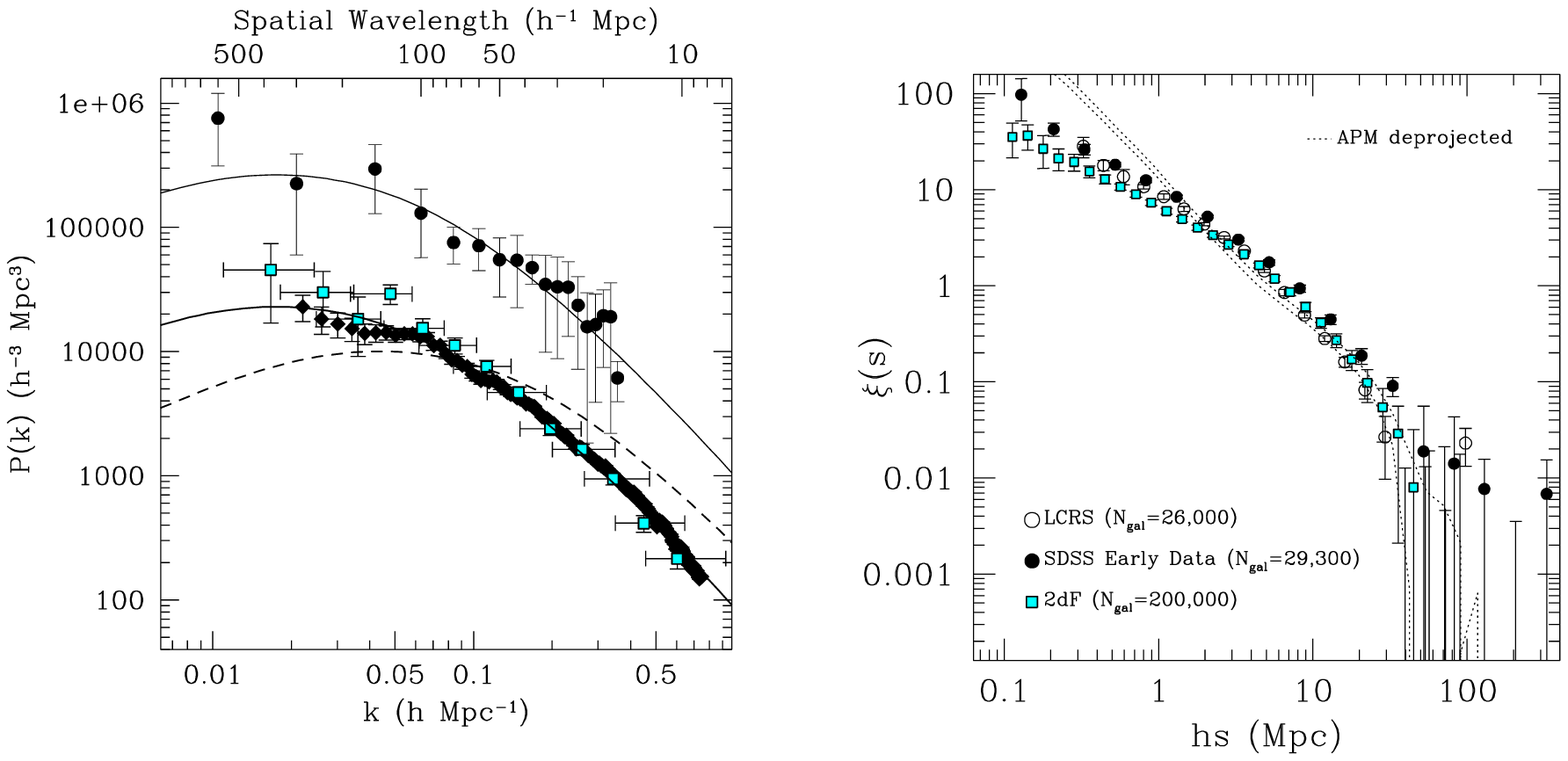}}

    \end{tabular}
   \end{center}
   \caption[pkrecent]
   { \label{fig:pkrecent}
The power spectrum of the 2dF galaxy redshift survey calculated
using two different estimators\cite{2dfpower,Tegmark01} and for
the X-ray built survey of clusters of galaxies REFLEX (filled
circles)\cite{reflex}. The dashed line correspond to an Einstein-de Sitter
($\Omega_M=1,\Omega_\Lambda=0$) CDM model that is clearly
rejected by the observations. The $\Lambda$-CDM concordance model
($\Omega_M=0.3,\Omega_\Lambda=0.7, h=0.7$) is represented by solid
lines with the appropriate bias parameters for both galaxies
(bottom line) and clusters (top line), (figure from
Guzzo\cite{guzzo02}.)}
   \end{figure}

The deeper the galaxy sample, the larger the spectral resolution
and the larger the wavenumber interval where the power spectrum
can be estimated. Fig.~\ref{fig:pkrecent} shows the power spectrum
for the 2dF survey when contained 160,000 galaxies and had a
depth of 750~$h^{-1}\,$Mpc. The power spectrum was calculated by
the 2dF team (filled diamonds) using the direct
method\cite{2dfpower}. The covariance matrix of this power
spectrum estimate was found from simulations of a matching
Gaussian Cox process in the sample volume. The diagram shows also
the results of the calculation performed by Tegmark {\it et
al.}\cite{Tegmark01} over the first public release of the sample
containing 102,000 redshifts (squares). Their paper is a 
good example of application of the large dataset machinery.
They used compression of the raw data into (pseudo) Karhunen-Lo\'eve 
eigenmodes and compressed these quadratically into band-powers,
using the Fisher matrix formalism to obtain the final
estimate of the power spectrum\cite{sgd,future}. 
For comparison, the power spectrum of the REFLEX cluster sample\cite{reflex}
(obtained by a direct method) is also shown.
Clusters of galaxies form only at the highest peaks of the density field.
This bias is the responsible for the larger amplitude of the power
spectrum corresponding to the clusters of galaxies. The diagram
also shows the curves for the power spectrum corresponding to
different models of structure formation (see the caption for the
details).

The main new feature in the spectra, obtained for the new deep
samples, is the emergence of details (wiggles). 
Sometime ago, the goal was to estimate the overall
behavior of the spectrum and, at most, to find its maximum, which
is related with the homogeneity scale. The new data enables us to
see and study the details of the spectrum. These wiggles could be
interpreted as traces of acoustic oscillations in the
post-recombination power spectrum. Similar oscillations are
predicted for the cosmic microwave background radiation
fluctuation spectrum. It seems, however, that the apparent wiggles
detected in the 2dF power spectrum are an artifact due to the
window function and other measurement
technicalities\cite{2dfpower,Tegmark01,guzzo02}.

\section{OTHER CLUSTERING MEASURES}
The correlation function can be generalized to higher
order\cite{sgd,gazta,peebval}: The $N$-point correlation
functions. This allows to statistically characterize the galaxy
distribution with a hierarchy of quantities which progressively
provide us with more and more information about the clustering of
the point process. These measures, however, had been difficult to
derive with reliability from the galaxy catalogs. The new
generation of surveys will surely overcome this problem.

There are, nevertheless, other clustering measures which provide
complementary information to the second-order quantities describe
above. For example, the topology of the large-scale structure
measured by the genus statistic\cite{gott} provides information
about the phase correlations of the density fluctuations in
$k$-space. To obtain this quantity, first the point process has to
be smoothed by means of a kernel function with a given bandwidth.
The topological genus of a surface is the number of holes minus
the number of isolated regions plus 1. This quantity is calculated
for the isodensity surfaces of the smoothed data corresponding to
a given density threshold. The genus has an analytically
calculable expression for a Gaussian field\cite{adler}.

Minkowski functionals are very effective clustering measures commonly used in
stochastic geometry\cite{skm}. These
quantities are adequate to study the shape and connectivity
of a union of convex bodies. They can easily be adapted to 
point processes\cite{mecke94} by considering the covering of the point
field formed by sets $A_r = \cup_{i=1}^N B_r({\mathbf x}_i)$ where
$r$ is the diagnostic parameter, $\{ {\mathbf x}_i \}_{i=1}^N$
represents the galaxy positions, and $B_r({\mathbf x}_i)$ is a ball
of radius $r$ centered at point ${\mathbf x}_i$. Minkowski
functionals\cite{kercher97} are applied to sets $A_r$ as $r$
increases. In $\realR^3$ there are four functionals: the volume
$V$, the surface area $A$, the integral mean curvature $H$, and
the Euler-Poincar\'e characteristic $\chi$, related with the genus
of the boundary of $A_r$.

Several quantities based on distances to nearest neighbors have
been used in the cosmological literature. The empty space function
$F(r)$ is the distribution function of the distance between a
given random test particle in $\realR^3$ and its nearest galaxy.
It is related with the void probability function\cite{lachieze}
$P_0(r)$
---the probability that a ball of radius $r$ randomly placed
contains no galaxies--- by $F(r)=1-P_0(r)$. $G(r)$ is the
distribution function of the distance $r$ of a given galaxy to its
nearest neighbor. The quotient $J(r) = [1-G(r)]/[1-F(r)]$ has been
successfully applied to describe the spatial pattern interaction
in the galaxy distribution\cite{kerscher99}. Related with the
nearest neighbor distances, the minimal spanning tree is a
structure descriptor that has shown powerful capabilities to
reveal the clustering properties of different point
patterns\cite{mstbarrow,mart90,doromst}. The minimal spanning tree
(MST) is the unique network connecting the $N$ points of the
process with a route formed by $N-1$ edges, without closed loops
and having minimal total length (the total length is
the sum of the lengths of the edges). The frequency histograms of the
MST edge lengths can be used to analyze the galaxy distribution and to
compare it with the simulated models\cite{doromst}.

The use of wavelets and related integral transforms is an
extremely promising tool in the clustering analysis of 3-D
catalogs. Some of these techniques are introduced in other
contributions published in this volume\cite{uno,dos}.

\acknowledgments     
We are grateful to our collaborators Martin Snethlage and Dietrich Stoyan 
for common results on the correlation function of shifted Cox
processes.
We thank  Valerie de Lapparent, Luigi Guzzo, Jon Loveday, and 
John Peacock for kindly providing us
with some illustrations.
This work was supported by the Spanish MCyT project AYA2000-2045 and
by the Estonian Science Foundation under grant 2882.


\bibliography{report}   
\bibliographystyle{spiebib}   
\end{document}